\documentclass{article}
\usepackage{spconf,amsmath}
\usepackage{graphicx}
\usepackage{amsmath}
\usepackage{amssymb}
\usepackage{comment}
\usepackage{bm}
\usepackage{multirow}
\usepackage{enumerate}
\usepackage{setspace}
\newcommand{\bmit}[1]{{\mbox{\boldmath $#1$}}}
\usepackage{siunitx}
\usepackage{bbding}
\usepackage{hyperref}
\usepackage{setspace}

\newlength\savedwidth
\newcommand{\wcline}[1]{\noalign{\global\savedwidth\arrayrulewidth\global\arrayrulewidth 1.0pt} \cline{#1}
\noalign{\global\arrayrulewidth\savedwidth}}

\title{Environmental sound synthesis \\from vocal imitations and sound event labels}

\name{\vspace{-0.3mm}\begin{tabular}{c} 
    Yuki Okamoto$^1$, 
    Keisuke Imoto$^2$, 
    Shinnosuke Takamichi$^3$, \\
    Ryotaro Nagase$^1$, 
    Takahiro Fukumori$^1$, 
    Yoichi Yamashita$^1$
    \end{tabular}
    \thanks{This work was supported by JSPS KAKENHI Grants Number 22KJ3027, 22H03639, and 23K16908.}
}
\address{
    $^1$Ritsumeikan University, Japan.
    $^2$Doshisha University, Japan.
    $^3$The University of Tokyo, Japan.
}
\begin{document}
\ninept
\maketitle
\setlength{\abovedisplayskip}{3pt} 
\setlength{\belowdisplayskip}{2pt} 
\setlength{\tabcolsep}{0.8mm} 
%
\begin{abstract}
\vspace{-3pt}
One way of expressing an environmental sound is using vocal imitations, which involve the process of replicating or mimicking the rhythm and pitch of sounds by voice.
We can effectively express the features of environmental sounds, such as rhythm and pitch, using vocal imitations, which cannot be expressed by conventional input information, such as sound event labels, images, or texts, in an environmental sound synthesis model.
In this paper, we propose a framework for environmental sound synthesis from vocal imitations and sound event labels based on a framework of a vector quantized encoder and the Tacotron2 decoder.
Using vocal imitations is expected to control the pitch and rhythm of the synthesized sound, which only sound event labels cannot control.
Our objective and subjective experimental results show that vocal imitations effectively control the pitch and rhythm of synthesized sounds. 
\end{abstract}

\vspace{-3pt}
\section{Introduction}
\vspace{-3pt}
\label{sec:intro}
Environmental sounds are effective for media content, such as video games and movies, to make them immersive and realistic.
Beyond the era of selecting environmental sounds from databases \cite{Gemmeke_ICASSP_2017,Fonseca_TASLP_2021}, generative approaches to environmental sounds based on deep learning are nowadays' the mainstream \cite{Kong_ICASSP2019,Zhou_CVPR_2018,Chen_TIP_2020}.
Environmental sound synthesis has many potential applications, such as the creation of background and sound effects for media content \cite{Zhou_CVPR_2018,Lloyd_ACMI3DGG_01} and data augmentation for environmental sound analysis \cite{Kong_ICASSP2019,Lloyd_ACMI3DGG_01,Gontier_ICASSP_2020,Salamon_WASPAA2017_01}.

Environmental sound synthesis methods using various types of input information have been proposed.
Table \ref{table:input_information} shows the list of the components of synthesized sounds controllable by the input information.
A method using sound event labels \cite{Kong_ICASSP2019,Liu_MLSP_2021} or images \cite{Zhou_CVPR_2018,Owens_CVPR_2016} as input can control the overall impression of sounds.
Other methods use textual information, e.g., onomatopoeic words (e.g., ``k a N k a N'')~\cite{okamoto_ATSIP_2022,ohnaka_arXiv_2022} and text prompts (e.g., ``Dogs bark behind the sound of people talking'')~\cite{Yang_arXiv_2022,Kreuk_ICLR_2023,Liu_arXiv_2023}.
These enable us to control the temporal changes and overall impression of the synthesized sounds.

However, these conventional methods used to represent sounds cannot express the pitch and rhythm of sounds.
As one way of expressing the pitch and rhythm of an environmental sound, we can use vocal imitations, which involve the process of replicating or mimicking the rhythm and pitch of sounds by voice.
We can express the target environmental sound characteristics by mapping its pitch, timbre, and rhythm to the properties of the voice \cite{Cartwright_CHI_2015}.
Vocal imitations are useful for expressing the characteristics of environmental sounds and are used in environmental sound retrieval \cite{Kim_DCASE_2018,Zhang_TASLP_2018,Cartwright_ICM_2014} from the audio database.

\begin{table}[t!]
\caption{Characteristics of synthesized environmental sound controllable by the input information}
\label{table:input_information}
\centering
\begin{tabular}{@{}l|ccccc@{}}
    \wcline{1-5} & \\[-8pt]
    \multirow{2}{*}{System input} & \multirow{2}{*}{Pitch} & \multirow{2}{*}{Rhythm} & Temporal & Overall \\
    & & & change & impression\\
    \hline \hline
    & \\[-8pt]
    Sound event label& & & & \Checkmark \\
    \cline{1-5}
    Image & & &  & \Checkmark \\
    \cline{1-5}
    Onomatopoeic word & & & \Checkmark &\\
    \cline{1-5}
    Caption & & & \Checkmark & \Checkmark \\
    \cline{1-5}
    Vocal imitation & \Checkmark &  \Checkmark &  \Checkmark &\\
    \wcline{1-5}
\end{tabular}%
\end{table}
\begin{figure}[t!]
\centering
\includegraphics[scale=0.75]{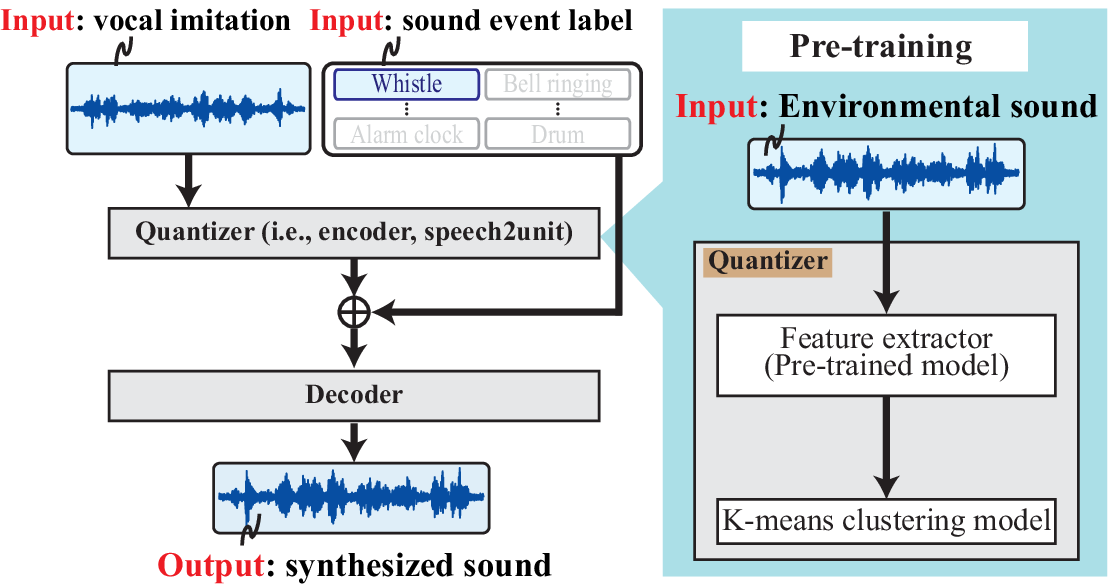}
\vspace{-6pt}
\caption{Proposed method. An environmental sound is synthesized from a vocal imitation and a sound event label.}
\label{fig:overview}
\end{figure}

In this paper, we propose a method of environmental sound synthesis from vocal imitations and sound event labels, as shown in Figure~\ref{fig:overview}.
Our method benefits from the characteristics listed in Table 1.
The use of vocal imitations enables us to control pitch and rhythm, and the use of sound event labels enables us to control the overall impression.
An encoder of our model is pre-trained on environmental sounds and quantizes the input vocal imitation; this makes the robust synthesis of environmental sounds.
Experimental evaluations clarify that our method can synthesize high-quality sounds and provide controllability of the pitch and rhythm of synthesized sounds.

\vspace{-3pt}
\section{Related work}
\vspace{-3pt}
\label{sec:related_work}
Takizawa et al. proposed a method of explosion sound synthesis from vocal imitation \cite{Takizawa_IUI_2023}.
This method enables us to control the nuance of explosion sounds by vocal imitation.
However, this method only targets the explosion sound category and cannot be extended to multiple categories. 
This is because vocal imitation is inherently one-to-many mapping, e.g., humans imitate the sound of the categories ``gunshot'' and ``balloons breaking'' in the same voice.
Therefore, only vocal imitation does not provide sufficient conditioning for sound synthesis.

On the other hand, one way to change the category of environmental sounds is to use sound event labels \cite{Kong_ICASSP2019,Liu_MLSP_2021}.
As explained in Section~\ref{sec:intro}, the fine control of environmental sounds is difficult using only sound event labels. 
However, it is expected that the disadvantages of both methods can be compensated by combining them with the aforementioned vocal imitation.

In a conventional work, the frame-level continuous representation extracted using the Transformer encoder has been used \cite{Takizawa_IUI_2023}.
However, in such cases, if the input vocal imitation fluctuates owing to noise or other factors, the noise of the vocal imitation may affect the quality of the synthesized sound.
To reduce the effect of the degradation of the input, the method of using the frame-level discrete representation has been proposed in speech synthesis and reconstruction \cite{Park_INTERSPEECH_2023}.
In environmental sound synthesis, reducing the effect of the degradation of the input can also be expected by using the discrete representation.

\vspace{-2pt}
\section{Proposed method}
\vspace{-2pt}
\label{sec:proposed_method}
Our model $\mathsf{Synthesizer}(\cdot)$ synthesizes environmental sound $\hat{{\bmit y}}$ from a vocal imitation ${\bmit x}$ and one-hot sound event label ${\bmit c}$:
\begin{align}
    \hat{{\bmit y}} = \mathsf{Synthesizer}({\bmit x}, {\bmit c}).
    \label{eq:train}
\end{align}
The model parameters are estimated using pairs of the ground-truth environmental sound ${\bmit y}$ and $\{{\bmit x}, {\bmit c}\}$.

Figure \ref{fig:overview} shows the model architecture that consists of an encoder and a decoder.
The encoder consists of a feature extractor and a $k$-means clustering model; it extracts quantized acoustic feature vectors from input vocal imitations.
We used pre-trained Bootstrap Your Own Latent for Audio (BYOL-A) \cite{Daisuke_TASLP_2023} as the feature extractor.
The continuous feature vectors extracted by BYOL-A are quantized by the $k$-means clustering model.
As explained in Section~\ref{sec:related_work}, quantized features are expected to reduce the effect of the degradation of input vocal imitation.
We pre-trained the k-means clustering model using environmental sounds on the ESC-50 dataset \cite{Piczak_CoM2_015}, which maps each frame of the feature of vocal imitation to the corresponding environmental sound features.
The conditioned discrete representations $\bm{V'}$ are calculated as 
\begin{equation}
\bm{V'}=\mathsf{Linear}\left(\vphantom{\bm{V},\left[\vphantom{\bm{c},\bm{c},\dots,\bm{c}}\right.}\right.
\bm{V},[\underbrace{\bm{c},\bm{c},\dots,\bm{c}}_{T}]
\left.\vphantom{\bm{V},\left.\vphantom{\bm{c},\bm{c},\dots,\bm{c}}\right]}\right)\in\mathbb{R}^{D\times T},
\end{equation}
where $\bm{V}$ and $\mathsf{Linear}(\cdot)$ denote the $T$-length $D$-dimensional discrete representations obtained by the encoder and linear transformation, respectively.
The quantized features control the pitch and rhythm of the synthesized sound, and the sound event label controls the overall impression.
The decoder that outputs mel-spectrograms follows the Tacotron2 \cite{Shen_ICASSP_2018} paper. 
The linear layer and decoder are trained to minimize the mean squared error between the ground truth and predicted mel-spectrograms.
The sound waveform is synthesized from the predicted mel-spectrogram using a neutral vocoder.

\vspace{-3pt}
\section{Experiments}
\vspace{-3pt}
\label{sec:experiments}
\subsection{Experimental setup}
\vspace{-3pt}
\textbf{Recording of vocal imitations.}
Because there are no suitable datasets for our purpose, we newly recorded vocal imitations corresponding to the environmental sounds of the ESC-50 dataset \cite{Piczak_CoM2_015}.
We presented an environmental sound to human imitators.
The imitators listened to the sound and then uttered it to imitate the sound. 
Imitators were allowed to listen and record again as many times as they wanted. 

Some environmental sounds in the ESC-50 dataset are diffiult to imitate with the human voice, such as multiple sounds appearing simultaneously. 
Therefore, we excluded sound events that include such difficult-to-imitate sounds and chose 31 sound event classes.
Each class included  $40$ audio samples, and we recorded $1,240$ ($40$ samples $\times$ $31$ sound events) vocal imitation samples. 
We hired three males and three females as the human imitators.
We collected a total of $7,440$ samples ($31$ sound event classes $\times$ $40$ samples $\times$ $6$ imitators)  of vocal imitations and divided the imitators into four (two males and two females) for training and two for evaluation.
We used $35$ samples per class of the four imitators for training and $5$ samples per class of the two imitators for evaluation.

We used a SHURE MX150B O-XLR microphone and a Roland Rubix24 audio interface for recording.
We saved the vocal imitations in the \SI{48}{kHz}, $16$ bit Linear PCM format.
\\\textbf{Model and learning configuration.}
The length, sampling rate, and waveform encoding of each environmental sound were $5$ seconds, $22.05$ kHz, and $16$-bit linear PCM, respectively.
We used a Hamming window with a length of $1,024$ samples and a hop length of $256$ samples, and obtained $80$-dimensional mel-spectrograms.
The model configurations of BYOL-A followed the open-sources implementation\footnote{\scriptsize{\url{https://github.com/nttcslab/byol-a/tree/master/v2}}}.
The $k$-means clustering model for the encoder was trained with $200$ clusters and $100$ iterations.
The number of clusters was set in accordance with the other speech and acoustic synthesis tasks utilizing vector quantization-based methods \cite{Lakhotia_TACL_2021}.
We used the one-hot encoded $31$-dimensional sound event labels.
The number of layers and units of LSTM of the decoder were $2$ and $1024$, respectively.
The number of pre-net layers and hidden dimensions of the decoder were $2$ and $256$, respectively.
The RAdam \cite{Liu_ICLR_2020} optimizer was used for training with a learning rate of $1.0 \times 10^{-4}$.
The batch size was set to $64$.
We used a neural vocoder\footnote{\scriptsize{\url{https://github.com/liuxubo717/sound_generation}}} pre-trained on an UrbanSound8K dataset \cite{Salamon_ACM-MM_2014}.
\vspace{-3pt}
\subsection{Evaluations}
\vspace{-3pt}
The synthesized sound should be as natural as an environmental sound.
To evaluate whether the proposed method is natural and can control pitch and rhythm, we conducted both subjective and objective evaluations in this study.
In Section~\ref{eval:sound_quality}, we evaluate the naturalness of the synthesized sounds.
In Sections~\ref{eval:sound_appropriateness_pitch} and \ref{eval:sound_appropriateness_rhythm}, we determine whether the synthesized sounds appropriately reflect the pitch and rhythm of the input vocal imitation, respectively.
In Sections~\ref{eval:pitch_shift} and \ref{eval:time_shift}, we evaluate whether the method follows changes in pitch and rhythm of vocal imitation, respectively.
In each subjective evaluation, we assigned ten listeners per sound.

We used three synthesis methods for evaluation: only sound event labels (``Label''), our method (``Label+vocal''), and analysis-synthesized sounds by the neural vocoder (``Reconstructed'').
We used the baseline model of DCASE 2023 challenge task 7\footnote{\scriptsize{\url{https://github.com/DCASE2023-Task7-Foley-Sound-Synthesis/dcase2023_task7_baseline}}} for ``Label.''

%

\begin{table}[t!]
\caption{MOS and standard deviation of sound naturalness in each sound event class $(\uparrow)$. Items in {\bf bold} indicate a significant difference between ``Label'' and ``Label+vocal''}
\label{table:result_quality}
\centering
\footnotesize
\resizebox{\linewidth}{!}{%
\begin{tabular}{@{}lrrrr@{}}
    \wcline{1-5}
     & \\[-8pt]
     \multirow{2}{*}{Sound event} & can & \multirow{2}{*}{fireworks} & door wood & {\bf average}\\
    & opening &  & knock & (all events)\\
    \hline \hline
    & \\[-8pt]
    Reconstructed & $3.88 \pm 1.04$ & $3.64 \pm 1.05$ & $3.90 \pm 0.97$ & $3.80 \pm 1.07$\\
    \cline{1-5}
    Label & $3.00 \pm 1.54$ & $2.90 \pm 1.42$ & $3.42 \pm 1.62$ & $\mathbf{2.52 \pm 1.25}$\\
    \cline{1-5}
    Label+vocal {\bf (proposed)} & $3.20 \pm 1.34$ & $\mathbf{3.66 \pm 1.14}$ & $3.92 \pm 1.03$ & $2.36 \pm 1.04$\\
    \wcline{1-5}
\end{tabular}%
\vspace{-10pt}
}
\end{table}
%
%

%
\begin{table}[t!]
\caption{MOS and standard deviation of sound appropriateness of pitch for input vocal imitations in each sound event class $(\uparrow)$. Items in {\bf bold} indicate a significant difference between ``Label'' and ``Label+vocal''}
\label{table:result_appropriateness_pitch}
\centering
\footnotesize
\resizebox{\linewidth}{!}{%
\begin{tabular}{@{}lrrrr@{}}
    \wcline{1-5}
     & \\[-8pt]
     \multirow{2}{*}{Sound event} & can & \multirow{2}{*}{fireworks} & door wood & {\bf average}\\
    & opening &  & knock & (all events)\\
    \hline \hline
    & \\[-8pt]
    Reconstructed & $3.82 \pm 1.00$ & $3.50 \pm 1.20$ & $3.90 \pm 1.05$ & $3.81 \pm 1.03$\\
    \cline{1-5}
    Label & $2.92 \pm 1.19$ & $2.66 \pm 1.21$ & $2.88 \pm 1.19$ & $2.54 \pm 1.15$\\
    \cline{1-5}
    Label+vocal {\bf (proposed)} & $\mathbf{3.78 \pm 1.04}$ & $3.10 \pm 0.97$ & $\mathbf{3.44 \pm 0.91}$ & $\mathbf{2.65 \pm 1.04}$\\
    \wcline{1-5}
\end{tabular}%
}
\end{table}

\vspace{-3pt}
\subsubsection{Sound naturalness}
\vspace{-3pt}
\label{eval:sound_quality}
To evaluate the naturalness of the synthesized sounds, we conducted the mean opinion score (MOS) test for naturalness.
We presented a sound and a sound event label to listeners. The listeners scored the naturalness of the sound on a scale of $1$ (very unnatural) to $5$ (very natural).

Table \ref{table:result_quality} shows scores of some event labels.
``average'' means averaged scores over all the 31 labels.
The results show that the method using voice imitation and sound event labels scored slightly lower in overall sound naturalness than the method using only sound event labels.

Figure~\ref{fig:MOS_distributions} (a) shows the MOS distributions of ``Label + vocal'' and ``Label''.
From the statistical significance test, the results in $5$ out of $31$ sound events also showed that our method provided higher sound naturalness than the label-based method.
Moreover, since there is no statistically significant difference between the methods in the results in 11 out of 31 sound events, we can consider that our proposed method has similar naturalness for nearly half of the sound events to the conventional method.
In addition, ``fireworks'' and ``door wood knock'' sounds generated by the proposed method are equivalent to the reconstructed sounds, whereas those generated by the label-based method did not achieve such naturalness.
These results show that the proposed method can generate sounds with relatively high naturalness for temporally sparse sounds. 
\vspace{-3pt}
\subsubsection{Sound appropriateness for pitch of input vocal imitations}
\vspace{-3pt}
\label{eval:sound_appropriateness_pitch}
We conducted a subjective evaluation to determine whether the synthesized sound was appropriate as a sound that reflected the pitch of the input vocal imitation.
We presented a synthesized sound, a vocal imitation, and a sound event label to listeners.
In this evaluation, we instructed the listeners to evaluate how well the synthesized sounds expressed the pitch of input vocal imitations.
The listeners scored the appropriateness of the sound for a given vocal imitation and sound event label on a scale of $1$ (not very appropriate) to $5$ (very appropriate).

Table \ref{table:result_appropriateness_pitch} shows the results.
Focusing on the average score, since there is a statistically significant difference between ``label'' and ``label+vocal'', we can say that the method using vocal imitations and sound event labels can generate sounds that best represent the pitch of input vocal imitations compared with the method using only sound event labels.

Figure~\ref{fig:MOS_distributions} (b) shows the MOS distribution of ``Label + vocal'' and ``Label''.
From the statistical significance test, the results in 6 out of 31 sound events also show that the method of using vocal imitations and sound event labels provided higher appropriateness than the method of using only sound event labels.
For animal sounds such as ``rooster'' and ``frog'', and temporally sparse sounds such as ``mouse click'', the proposed method was significantly different from the conventional method.

\vspace{-3pt}
\subsubsection{Sound appropriateness for rhythm of input vocal imitations}
\vspace{-3pt}
\label{eval:sound_appropriateness_rhythm}
We conducted a subjective evaluation to determine whether the synthesized sound was appropriate as a sound that reflected the rhythm of the input vocal imitation.
We conducted a subjective evaluation as in Section~\ref{eval:sound_appropriateness_pitch}.
In this evaluation, the appropriateness is evaluated for rhythm, not pitch.

Table \ref{table:result_appropriateness_rhythm} shows the results.
Focusing on the average score, since there is a statistically significant difference between ``label'' and ``label+vocal'', we can say that the method using vocal imitations and sound event labels can generate sounds that best represent the rhythm of input vocal imitations compared with the method using only sound event labels.

\begin{figure*}[t]
\centering
\includegraphics[width=0.85\linewidth]{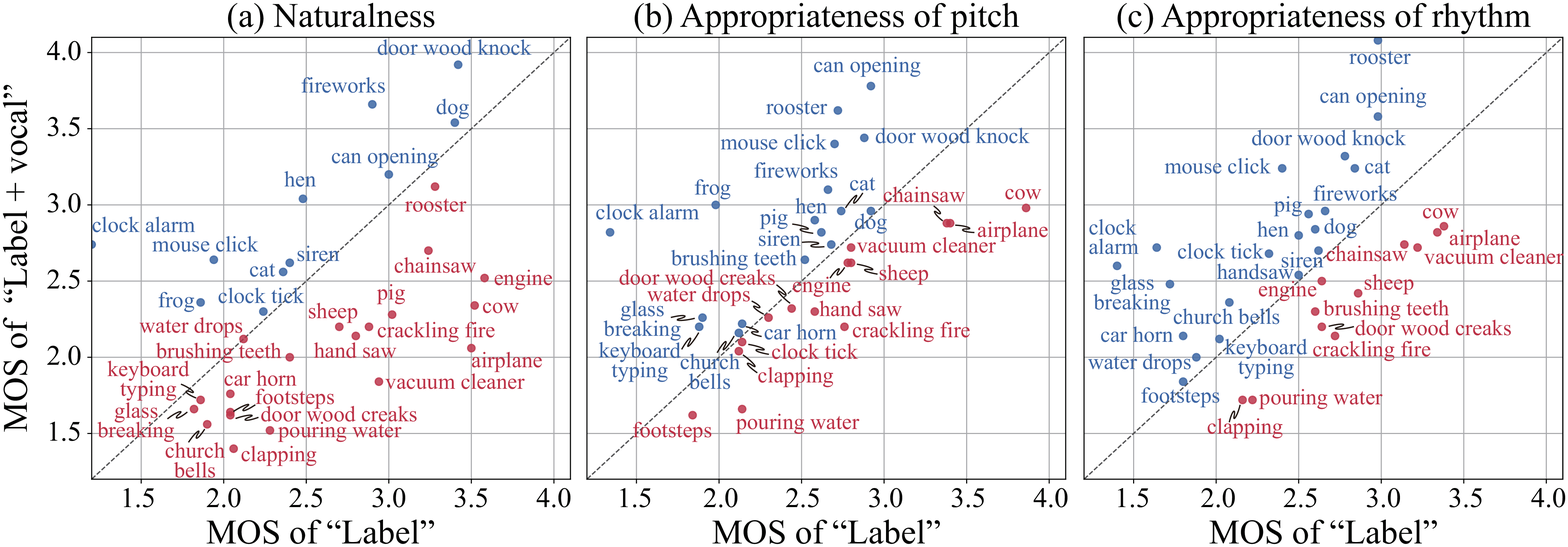}
\vspace{-8pt}
\caption{MOS distributions: ``Label'' vs ``Label + vocal.'' Blue points mean that ``Label + vocal.'' is superior to ``Label.''}
\label{fig:MOS_distributions}
\vspace{-6pt}
\end{figure*}

Figure~\ref{fig:MOS_distributions} (c) shows the MOS distribution of ``Label + vocal'' and ``Label''.
From the statistical significance test, the results in $7$ out of $31$ sound events show that the method of using vocal imitations and sound event labels provided higher appropriateness than the method of using only sound event labels.
As in Section~\ref{eval:sound_appropriateness_pitch}, for animal sounds and temporally sparse sounds, the proposed method was significantly different from the conventional method.

In the evaluations in Sections~\ref{eval:sound_appropriateness_pitch} and \ref{eval:sound_appropriateness_rhythm}, intermittent sounds such as ``engine'' tended to score lower.
In the future, further analysis of sounds we did not synthesize well, such as ``engine,'' should be carried out.

\begin{table}[t!]
\caption{MOS and standard deviation of sound appropriateness of rhythm for input vocal imitations in each sound event class $(\uparrow)$. Items in {\bf bold} indicate a significant difference between ``Label'' and ``Label+vocal''}
\label{table:result_appropriateness_rhythm}
\centering
\footnotesize
\resizebox{\linewidth}{!}{%
\begin{tabular}{@{}lrrrr@{}}
    \wcline{1-5}
     & \\[-8pt]
     \multirow{2}{*}{Sound event} & can & \multirow{2}{*}{fireworks} & door wood & {\bf average}\\
    & opening &  & knock & (all events)\\
    \hline \hline
    & \\[-8pt]
    Reconstructed & $4.10 \pm 0.86$ & $3.68 \pm 0.79$ & $4.08 \pm 0.99$ & $3.87 \pm 1.01$\\
    \cline{1-5}
    Label & $2.98 \pm 1.19$ & $2.99 \pm 0.94$ & $2.78 \pm 1.33$ & $2.54 \pm 1.15$\\
    \cline{1-5}
    Label+vocal {\bf (proposed)} & $\mathbf{3.58 \pm 1.11}$ & $2.96 \pm 1.01$ & $\mathbf{3.32 \pm 1.19}$ & $\mathbf{2.62 \pm 1.09}$\\
    \wcline{1-5}
\end{tabular}%
}
\end{table}

\begin{figure}[t!]
\centering
\includegraphics[scale=0.88]{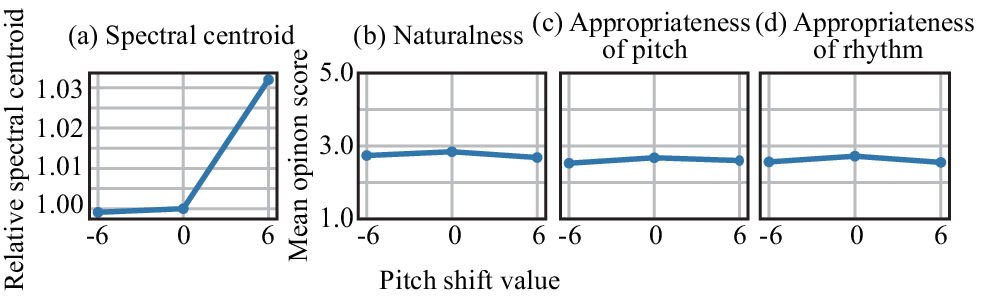}
\vspace{-18pt}
\caption{Results of objective and subjective evaluations of sound control by pitch change}
\label{fig:pitch_change}
\vspace{-10pt}
\end{figure}
\begin{figure}[t!]
\centering
\includegraphics[scale=0.88]{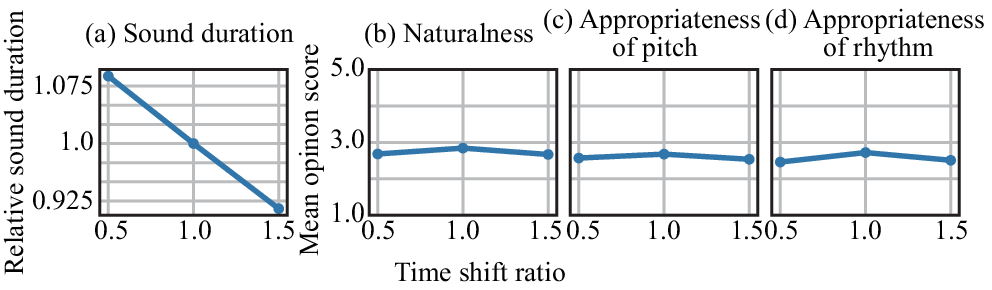}
\vspace{-18pt}
\caption{Results of objective and subjective evaluations of sound control by rhythm change}
\label{fig:rhythm_change}
\vspace{-10pt}
\end{figure}
\vspace{-3pt}
\subsubsection{Sound control by pitch change}
\vspace{-3pt}
\label{eval:pitch_shift}
We evaluated the capability to control the synthesized sound by changing the pitch of vocal imitation.
The pitch-changed vocal imitations were artificially generated by signal-processing-based pitch shifting by two values (in semi-octave): $\{-6, 6\}$.
For signal-processing-based pitch shifting, we used the function in {\it librosa}\footnote{\scriptsize{\url{https://librosa.org/doc/main/generated/librosa.effects.pitch_shift.html}}}.
The relative spectral centroid of the sound was objectively evaluated, setting the one synthesized from the original (i.e., unshifted pitch) vocal imitation to $1$.
The subjective evaluation was conducted in the same manner as described in Sections~\ref{eval:sound_quality}, \ref{eval:sound_appropriateness_pitch}, and \ref{eval:sound_appropriateness_rhythm}.

Figure \ref{fig:pitch_change} shows the results.
From Figure~\ref{fig:pitch_change} (a), the spectral centroid changes accordingly as pitch increases, so we can say that the pitch of the generated sound can be controlled by changing the pitch of the input vocal imitation.
On the other hand, as shown in Figures~\ref{fig:pitch_change} (b)--(d), changing the pitch slightly degraded each MOS score.
In the future,  we will investigate the reason for this.

\vspace{-3pt}
\subsubsection{Sound control by rhythm change}
\vspace{-3pt}
\label{eval:time_shift}
We evaluated the capability to control the synthesized sound by changing the rhythm of vocal imitation.
The rhythm-changed vocal imitations were artificially generated by signal-processing-based speed shifting by two ratios: $\{0.5, 1.5\}$.
For signal-processing-based speed shifting, we used the function in {\it librosa}\footnote{\scriptsize{\url{https://librosa.org/doc/main/generated/librosa.effects.time_stretch.html}}}.
The relative duration of the sound was objectively evaluated, setting the one synthesized from the original (i.e., unshifted speed) vocal imitation to $1$.
The subjective evaluation was conducted in the same manner as described in Sections~\ref{eval:sound_quality}, \ref{eval:sound_appropriateness_pitch}, and \ref{eval:sound_appropriateness_rhythm}

Figure \ref{fig:rhythm_change} shows the results.
From Figure~\ref{fig:rhythm_change} (a), the sound duration changes accordingly as the speed shift ratio increases, so we can say that the rhythm of the generated sound can be controlled by changing the rhythm of the input vocal imitation.
On the other hand, as shown in Figures~\ref{fig:rhythm_change} (b)--(d), changing the rhythm slightly degraded for each MOS score.
In the future,  we will investigate the reason for this.

\vspace{-3pt}
\subsubsection{Spectrograms of synthesized sounds}
\begin{figure}[t!]
\centering
\includegraphics[scale=0.89]{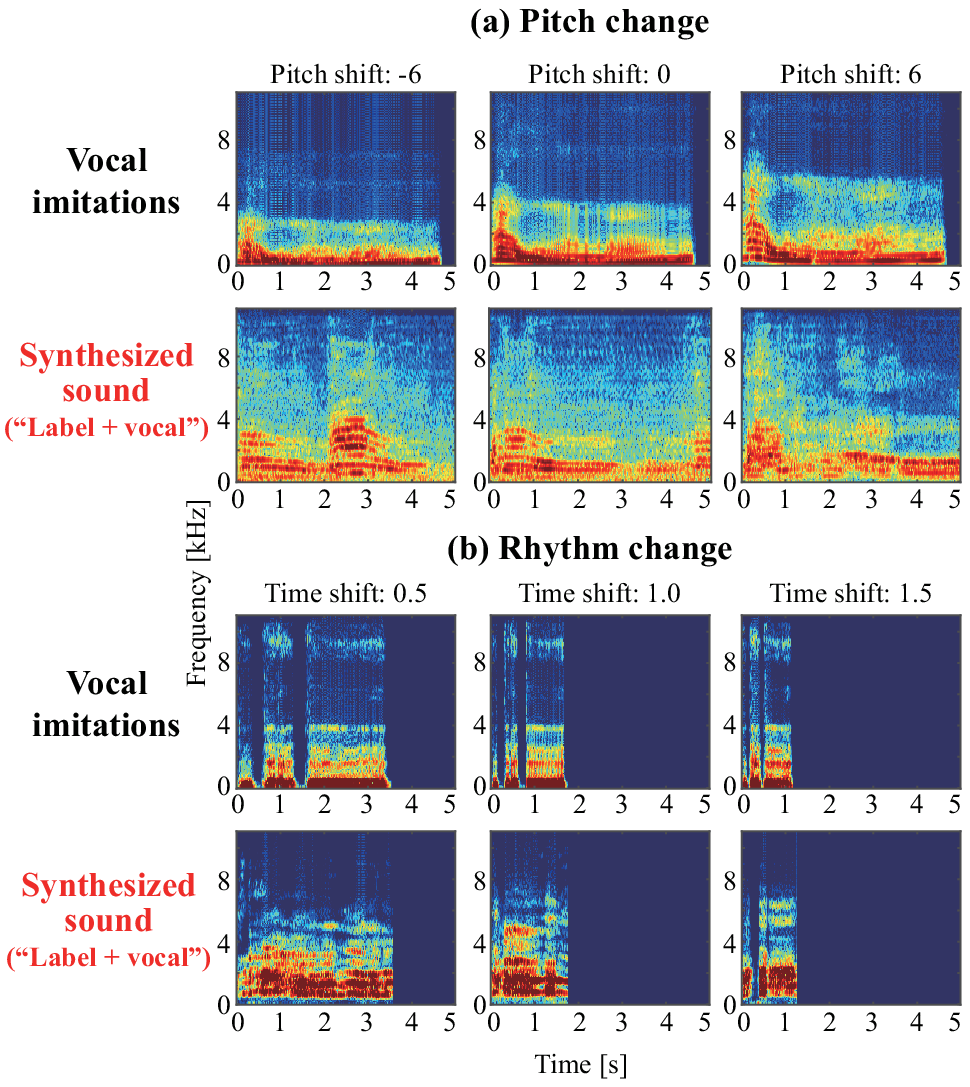}
\vspace{-20pt}
\caption{Spectrograms of pitch- and rhythm-changed vocal imitation and synthesized sounds}
\label{fig:spectro_result}
\vspace{-10pt}
\end{figure}

\vspace{-3pt}
\label{eval:spectro}
Figure \ref{fig:spectro_result} shows the spectrograms of the pitch- and rhythm-changed vocal imitation and synthesized sounds.
From this figure, we can confirm that the method using vocal imitations and sound event labels as inputs successfully reproduced the rhythm and pitch of vocal imitations, which were used as the input.
Thus, we consider that the use of vocal imitations, in addition to sound event labels, is effective in controlling the pitch and rhythm of synthesized sounds.
The synthesized sounds of each synthesis method are available on our web page\footnote{\scriptsize{\url{https://voice-to-foley.github.io/}}}.

\vspace{-6pt}
\section{Conclusion}
\label{sec:conclusion}
\vspace{-3pt}
In this paper, we proposed a method of environmental sound synthesis from vocal imitations and sound event labels.
Our experiments confirmed that our method can synthesize high-quality sounds that reflect the pitch and rhythm of the input vocal imitation.
Future work includes combining additional information, e.g., images.

\bibliographystyle{IEEEbib}
\bibliography{refs}

\end{document}